# Multidimensional Generalized Quadrature Index Modulation for 5G Wireless Communications

Taissir Y. Elganimi* and Khaled M. Rabie[Δ]
*Department of Electrical and Electronic Engineering, University of Tripoli, Libya
[Δ]Department of Engineering, Manchester Metropolitan University, United Kingdom
Emails: t.elganimi@uot.edu.ly and k.rabie@mmu.ac.uk

*Abstract*—Multidimensional generalized quadrature index modulation scheme is proposed in this paper for conveying extra digital information with the aid of the space, radio frequency (RF) mirrors, and time indices. Explicitly, this proposed scheme cleverly combines another proposed time-indexed generalized quadrature spatial modulation (TI-GQSM) system with media-based modulation (MBM) transmission principle using RF mirrors, and it is referred to as TI-GQSM-MBM scheme. This scheme is attractive because of both the high data rate and the significant performance improvements that can be achieved. The system performance of the proposed schemes in terms of the bit error rate (BER) is evaluated and compared to the performance of the conventional schemes. Simulation results showed that a significant improvement is achieved by the TI-GQSM-MBM scheme as compared to that of TI-GQSM, time-indexed media-based modulation (TI-MBM) and the conventional generalized quadrature spatial modulation (GQSM) schemes for the same rate. It is also demonstrated that the proposed schemes are robust to channel estimation errors (CEEs) as compared to multidimensional generalized spatial modulation (GSM) schemes. Therefore, the proposed schemes can be effectively used as an alternative solution for various 5G and beyond wireless networks.

## I. Introduction

Index modulation (IM) is an innovative technique in wireless communications that provides new dimensions to convey extra information by utilizing the indices of multiple transmission resources. For example, the indices of transmit antennas of multiple input multiple output (MIMO) systems in the spatial domain [1], the time slots in the time domain [2], the subcarriers in multicarrier systems [3], or the radio frequency (RF) mirrors [4, 5]. Several IM schemes and their significant performance improvement have attracted an increasing interest from researchers in the past decade, and have been regarded as promising solutions for the fifth generation (5G) and beyond wireless communication systems [6, and references therein].

The most popular IM transmission technique in the literature is the emerging spatial modulation (SM) technique [1] that utilizes transmit antenna indices for the transmission of extra digital information bits from single active antenna at each time instant. It has attracted considerable attention from researchers and considered as the pioneer of IM techniques because of its high data rate and low complexity. In [7], SM scheme is generalized to transmit the same signal symbol from multiple antennas $N_a$ out of total number of antennas $N_t$ simultaneously, and referred to as generalized SM (GSM). Later, another space modulation technique is proposed and referred to as quadrature SM (QSM) [8]. This technique expands the spatial constellation symbol of the conventional SM scheme into in-phase and quadrature components to include real and imaginary dimensions in order to transmit the real and imaginary parts of the constellation symbol. The recent studies of [9, 10] considered a generalized framework of QSM and referred to as generalized QSM (GQSM) scheme that combines the benefits of both GSM and QSM schemes.

Over the past few years, an emerging novel IM scheme has been carried out in the time domain, where in each data frame, a fraction of the time slots is activated for transmission, and the corresponding indices are utilized to carry information. This IM transmission technique is termed as space-time IM [2], and showed a significant performance improvement as compared to its non-IM-aided counterparts. In addition, IM transmission can be carried out in the channel domain. Specifically, media-based modulation (MBM) is a multi-antenna modulation scheme where each transmit antenna is equipped with digitally controlled parasitic elements that act as $m_{RF}$ mirrors to be used for signaling [4, 5]. These RF mirrors can be turned ON and OFF based on the incoming information bits in order to create $2^{m_{RF}}$ various complex channel fade realizations from the perspective of the receiver.

Recently, several research studies have focused on combining MIMO schemes with MBM transmission principle because of the fact that MBM signal vectors have good distance properties [11], which leads to achieving an improved performance in MBM-based schemes over the conventional modulation schemes. For instance, the authors in [12] and [13] have combined space-time block codes (STBC), and uncoded space-time labelling diversity schemes with MBM concept, respectively. In other research studies, both GSM and QSM techniques have been integrated with MBM approach as presented in [14] and [15], respectively. The scheme that proposed in [15] combines QSM system with MBM transmission technique, and termed as quadrature channel modulation (QCM).

Motivated by the performance enhancement that can be achieved through the use of index bits, combining multiple transmission resources as presented in [11] in the conventional SM scheme is beneficial. Therefore, this paper introduces the concept of *generalized quadrature channel modulation (GQCM)* by generalizing the QCM scheme and cleverly combining both GQSM and MBM principles. It is in order to further enhance the data rate of SM-based MBM scheme with reducing the number of transmit antennas. In addition, this paper proposes the time-indexed GQSM (TI-GQSM) and time-indexed GQCM that is referred to as *time-indexed generalized quadrature spatial modulation media-based modulation (TI-GQSM-MBM)*, where three entities, namely, antennas, time slots, and RF mirrors are utilized and indexed simultaneously. Thus, digital information bits are conveyed in the TI-GQSM-MBM scheme through indices of the used time slots, indices ($\ell_\Re$ and $\ell_\Im$) of the active antennas, indices of the RF mirrors, besides the conventional modulation bits (i.e., $\log_2 M$), where $M$ is the constellation size. Additionally, TI-GQSM and GQCM transmission schemes are introduced as special cases of the proposed TI-GQSM-MBM scheme. In this paper, the bit error rate (BER) performance and the rate analysis of the proposed schemes are evaluated and compared to time-indexed media-based modulation (TI-MBM) scheme where time slots and RF mirrors are indexed [16]. The simulation results suggest that the proposed TI-GQSM-MBM scheme can be a promising multi-antenna modulation scheme that has the potential for further investigations.

The rest of the present paper is organized as follows. Section II presents the system model of the proposed schemes. Section III shows the BER performance of the proposed schemes. Finally, conclusions are summarized in Section IV.

*Notation:* The upper- and lower-case boldface letters denote matrices and vectors, respectively. $\|\mathbf{A}\|$ denotes the Frobenius norm operation of $\mathbf{A}$, $\lfloor \cdot \rfloor$ stands for the floor operation that flooring a real number to the nearest smallest integer, $\binom{a}{b}$ denotes the combinations without repetition of $a$ objects taken $b$ at a time, and $\mathbb{C}^{m \times n}$ denotes a complex matrix having the size of $m \times n$.

## II. SYSTEM MODEL

In this paper, a new multidimensional generalized quadrature index modulation is proposed as depicted in Fig. 1, where information bits are conveyed through indexing of multiple entities simultaneously. More specifically, the following three schemes are proposed and considered:

### A. Time-Indexed Generalized Quadrature Spatial Modulation (TI-GQSM)

In this subsection, TI-GQSM scheme is considered, where time slots and antennas are indexed simultaneously. The TI-GQSM scheme has $N_t$ transmit antennas and $N_a < N_t$ active antennas, and the information bits are conveyed through indexing the time-slots and active antennas, besides quadrature amplitude modulation (QAM) or phase shift keying (PSK) symbols. In time-indexed schemes, the time is divided into multiple frames, and each frame consists of $T + L - 1$ time slots, where $T$ is the length of the corresponding data part of the frame in number of time slots, $L$ is the number of taps in the multipath frequency-selective channel, and $L - 1$ is the number of time slots that used for transmitting cyclic prefix (CP) [11]. Time indexing in the proposed TI-GQSM scheme is done by selecting only $T_a$ time slots out of $T$ time slots in a frame to convey $\left\lfloor \log_2 \binom{T}{T_a} \right\rfloor$ information bits, which are called 'time index bits'. The $T$-length pattern of active and inactive status of the time slots is called a time-slot activation pattern (TAP). In each active time slot, $N_a$ active antennas are activated to transmit both the real and imaginary parts of the signal symbol based on $2 \left\lfloor \log_2 \binom{N_t}{N_a} \right\rfloor$ bits. These bits are called 'antenna combinations index bits'. The active antennas in TI-GQSM scheme can be different in each active slot, as well as for conveying the real and imaginary parts. Since there are $T_a$ time slots out of $T$ signaling time slots in each frame, a number of $2T_a \left\lfloor \log_2 \binom{N_t}{N_a} \right\rfloor$ bits are transmitted through antenna indexing in each frame in TI-GQSM scheme. In addition, $T_a \log_2(M)$ information bits are conveyed by the conventional modulation symbols in each frame. As a result, the achieved rate of the proposed TI-GQSM scheme in bits per channel use (bpcu) is written as

$$\eta_{TI-GQSM} = \frac{1}{T+L-1} \left\{ \left\lfloor \log_2 \binom{T}{T_a} \right\rfloor \right.$$
$$\left. + T_a \left[ 2 \left\lfloor \log_2 \binom{N_t}{N_a} \right\rfloor + \log_2(M) \right] \right\}. \quad (1)$$

Fig. 1 specializes TI-GQSM scheme by removing RF mirror ON/OFF control switch and RF mirrors. Like in multidimensional SM scheme proposed in [11], an $N_t \times 1$ GQSM signal vector gets conveyed in each active time slot. After removing the CP, the received signal vector is written as

$$\mathbf{y} = \mathbf{H}\mathbf{s} + \mathbf{n} \quad \in \mathbb{C}^{TN_r \times 1}, \quad (2)$$

where $\mathbf{n} \in \mathbb{C}^{TN_r \times 1}$ denotes a complex zero-mean additive white Gaussian noise (AWGN) vector with a variance of its elements being $\sigma_n^2$ per dimension at the receiver input, $\mathbf{n} \sim \mathcal{CN}(\mathbf{0}_{N_r}, \sigma_n^2 \mathbf{I}_{N_r})$, $\mathbf{I}_{N_r}$ is the $N_r \times N_r$ identity matrix, and $\mathbf{H} \in \mathbb{C}^{TN_r \times TN_t}$ is the equivalent channel matrix that written as

$$\mathbf{H} = \begin{bmatrix} \mathbf{H}_0 & \mathbf{0} & \mathbf{0} & \cdots & \mathbf{H}_{T_a-1} & \cdots & \mathbf{H}_1 \\ \mathbf{H}_1 & \mathbf{H}_0 & \mathbf{0} & \cdots & \mathbf{0} & \cdots & \mathbf{H}_2 \\ \vdots & & & & \vdots & & \vdots \\ \mathbf{H}_{T_a-1} & \mathbf{H}_{T_a-2} & \vdots & \mathbf{H}_0 & \mathbf{0} & \cdots & \mathbf{0} \\ \mathbf{0} & \mathbf{H}_{T_a-1} & \vdots & \mathbf{H}_1 & \mathbf{H}_0 & \cdots & \mathbf{0} \\ \vdots & & & & \vdots & \ddots & \\ \mathbf{0} & \mathbf{0} & \cdots & \cdots & \cdots & & \mathbf{H}_0 \end{bmatrix}, \quad (3)$$

where $\mathbf{H}_i \in \mathbb{C}^{N_r \times N_t}$ is the channel matrix of the $i$-th active time slot, $i = 0, 1, \ldots, T_a - 1$. It is assumed that the $k$-th column of $\mathbf{H}_i$ has zero mean and unit variance per dimension, $\mathbf{h}_i^k \sim \mathcal{CN}\left(\mathbf{0}_{N_r}, \frac{1}{L} \mathbf{I}_{N_r}\right)$. Without loss of generality, a flat Rayleigh fading channel with single path ($L = 1$) is assumed in this paper.

### B. Generalized Quadrature Spatial Modulation Media-Based Modulation (GQSM-MBM)

The proposed GQSM-MBM scheme is based on combining both GQSM and MBM transmission schemes, and can be also referred to as GQCM. The advantage of integrating MBM in MIMO systems is that the number of information bits that conveyed by indexing the $m_{RF}$ mirrors increases linearly with the number of RF mirrors. Moreover, indexing in GQSM-MBM scheme that corresponds to exploiting the channel domain is done across MBM transmit unit (MBM-TU) which includes the transmit antenna and its surrounding $m_{RF}$ mirrors. In this scheme, there are $N_t$ MBM-TUs, but only $N_a$ MBM-TUs are selected and activated for transmitting each part (real or imaginary) of the signal symbol at each time instant based on $2 \left\lfloor \log_2 \binom{N_t}{N_a} \right\rfloor$ bits, and an $N_t \times 1$ GQSM signal vector is conveyed from $N_t$ MBM-TUs. Additionally, the $m_{RF}$ mirrors that associated with each active MBM-TU are made ON and OFF depending on the $m_{RF}$ bits that are chosen by the mirror activation pattern (MAP) [11]. In GQSM-MBM scheme, the channel states are exploited for transmitting additional bits besides the indexing bits of both the in-phase and quadrature components of the complex signal symbols, $2\left\lfloor\log_2\binom{N_t}{N_a}\right\rfloor$, and the conventional modulation bits, $\log_2(M)$. Therefore, the achieved rate of the proposed GQSM-MBM scheme (in bpcu) is expressed as

$$\eta_{GQSM-MBM} = \frac{\left[2\left\lfloor\log_2\binom{N_t}{N_a}\right\rfloor + m_{RF} + \log_2(M)\right]}{L}. \quad (4)$$

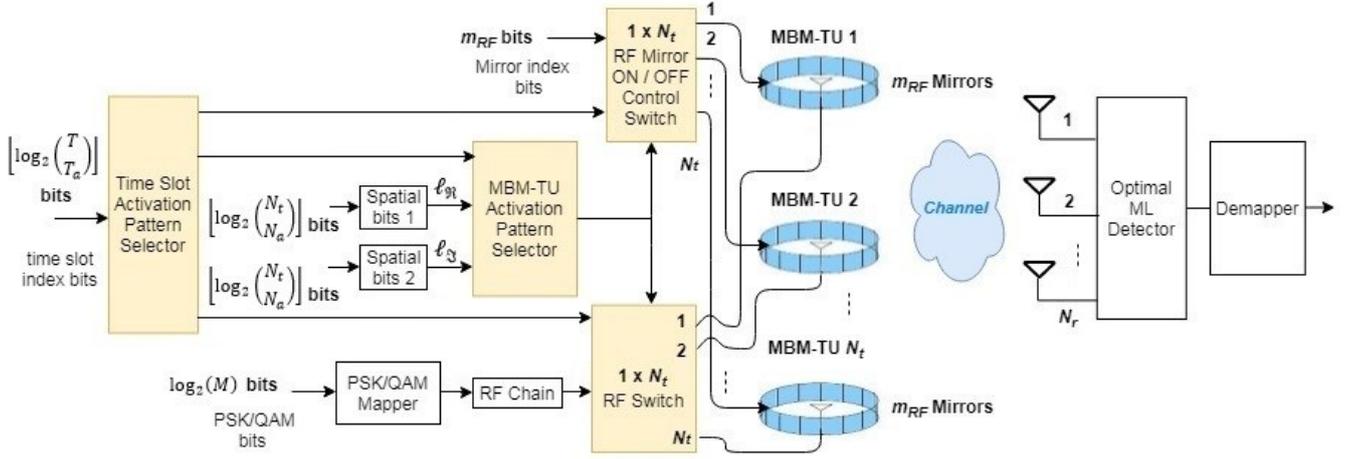

Fig. 1. System model of the proposed TI-GQSM-MBM scheme with single RF chain.

Fig. 1 can be specialized to GQSM-MBM scheme if the TAP selector is removed, as time indexing is not involved in GQSM-MBM scheme. In addition, (2) can be used in this transmission technique, where $\mathbf{H} \in \mathbb{C}^{N_r \times N_t 2^{m_{RF}}}$ is the channel matrix, and $\mathbf{n} \in \mathbb{C}^{N_r \times 1}$ is the noise vector.

### C. Time-Indexed GQSM-MBM (TI-GQSM-MBM)

In this paper, the proposed TI-GQSM-MBM scheme is considered where efficiently performs indexing in time, space and channel domains (i.e., in time slots, antennas and RF mirrors). This scheme has $N_t$ MBM-TUs, $N_a$ active transmit antennas and single RF chain, and indexing in time and channel domains occurs as explained earlier in TI-GQSM and GQSM-MBM schemes, respectively. Equations (2) and (3) can be used in this scheme, where $\mathbf{H}_i \in \mathbb{C}^{N_r \times N_t 2^{m_{RF}}}$ being the effective channel matrix corresponding to the $i$-th active time slot. TI-GQSM-MBM scheme is a generalized transmission scheme of which TI-GQSM, TI-MBM and GQSM-MBM schemes are special cases when $m_{RF} = 0$, $N_t = N_a = 1$ [16] and $T = T_a = 1$, respectively. Consequently, the achieved rate of the proposed TI-GQSM-MBM scheme (in bpcu) is expressed as

$$\eta_{TI-GQSM-MBM} = \frac{1}{T+L-1}\left\{\left\lfloor\log_2\binom{T}{T_a}\right\rfloor \right.$$
$$\left. + T_a\left[2\left\lfloor\log_2\binom{N_t}{N_a}\right\rfloor + m_{RF} + \log_2(M)\right]\right\}. \quad (5)$$

It is worth noting that TI-GSM-MBM scheme can be designed in the same manner of the proposed TI-GQSM-MBM scheme, except that the term of the indexing bits of both the in-phase and quadrature components of the complex signal symbols, $2\left\lfloor\log_2\binom{N_t}{N_a}\right\rfloor$ in (5) is modified and divided by 2, since there is no expanding for the spatial constellation symbols. Similarly, a new TI-GSM scheme can be considered as a special case of the proposed TI-GQSM scheme, which can be also designed by generalizing the TI-SM scheme presented in [2]. It is also noticeable that TI-SM-MBM scheme introduced in [11] and TI-QSM-MBM are special cases of the proposed TI-GSM-MBM and TI-GQSM-MBM transmission schemes when $N_a = 1$, respectively.

### D. Maximum Achieved Rate

Due to the presence of the binomial coefficient term $\binom{T}{T_a}$ in (5), it is clear that the achieved rate of TI-GQSM-MBM scheme increases with increasing $T_a$ up to an optimum value, $T_{opt}$, where it reaches its maximum value. Beyond this value, the rate decreases with increasing $T_a$. The optimum value of $T_a$ for which the rate is maximized is derived in this subsection with keeping $T$, $N_t$, $N_a$, $m_{RF}$, $M$ and $L$ fixed. If the term $2\left\lfloor\log_2\binom{N_t}{N_a}\right\rfloor + m_{RF} + \log_2(M)$ in (5) is defined as $\beta$, therefore, the achieved rate can be represented as

$$\eta_{TI-GQSM-MBM} = \frac{1}{T+L-1}\left\{\left\lfloor\log_2\binom{T}{T_a}\right\rfloor + T_a\beta\right\}. \quad (6)$$

According to [11], it is worth noting that

$$\log_2\binom{T}{T_a} + T_a\beta - 1 < \left\lfloor\log_2\binom{T}{T_a}\right\rfloor + T_a\beta$$
$$\leq \log_2\binom{T}{T_a} + T_a\beta. \quad (7)$$

It can be observed from (7) that the upper and lower bounds of $\left\lfloor\log_2\binom{T}{T_a}\right\rfloor + T_a\beta$ differ by a constant value of one for all values of $T_a$, where $1 \leq T_a \leq T$. Since the upper and lower bounds achieve their maximum value for the same value of $T_a$, and there can be only one integer between the two bounds in (7), only the upper bound is considered to find the value of $T_{opt}$ that maximizes the achieved rate of (5) [11]. According to [11] and [17], it can be found that

$$\frac{T_a}{T-T_a} = 2^\beta = 2^{\left(2\left\lfloor\log_2\binom{N_t}{N_a}\right\rfloor + m_{RF} + \log_2(M)\right)}. \quad (8)$$

Therefore, the optimum value of active time slots ($T_{opt}$) that achieves the maximum rate in TI-GQSM-MBM scheme can be found as

$$T_{opt} = T\left(\frac{2^\beta}{1+2^\beta}\right) = T\left(\frac{2^{\left(2\left\lfloor\log_2\binom{N_t}{N_a}\right\rfloor + m_{RF} + \log_2(M)\right)}}{1+2^{\left(2\left\lfloor\log_2\binom{N_t}{N_a}\right\rfloor + m_{RF} + \log_2(M)\right)}}\right). \quad (9)$$

Under specific system configurations, the proposed TI-GQSM-MBM scheme reduces to different IM schemes. Specifically, it is clear from (9) that it can be specialized to TI-GQSM and TI-MBM schemes as special cases of TI-GQSM-MBM scheme by substituting $m_{RF} = 0$ and $N_t = N_a = 1$ in (9), respectively. This equation can be also specialized to TI-GSM-MBM scheme with multiple active antennas and TI-QSM-MBM scheme, when $\beta$ equals to $\left(\left\lfloor\log_2\binom{N_t}{N_a}\right\rfloor + m_{RF} + \log_2(M)\right)$ and $\left(\log_2(N_t^2 M) + m_{RF}\right)$, respectively.

Fig. 2 depicts the achieved rate versus the number of active time slots $T_a \in \{1, 2, ..., T\}$ for TI-GQSM, TI-MBM, GQSM-MBM and TI-GQSM-MBM schemes with $T = 128$, $M = 2$ for binary PSK and $L = 1$. It can be clearly seen from this figure that the rate of TI-GQSM, TI-MBM and TI-GQSM-MBM schemes increases as $T_a$ increases until it reaches the maximum rate at a certain value of $T_a$ in each scheme. According to (9) and Fig. 2, the optimum value $T_{opt}$ that maximizes the achieved rate is approximately 113, 120 and 126 in TI-GQSM, TI-MBM and TI-GQSM-MBM schemes, respectively. It is also obvious from Fig. 2 that both GQSM-MBM and TI-GQSM-MBM techniques use the same values of $N_t$, $N_a$ and $m_{RF}$, and the same modulation scheme, $M = 2$, and the rate of GQSM-MBM is constant since the time indexing is not performed in this scheme. Additionally, the achieved rate of the proposed TI-GQSM-MBM scheme is higher than that of TI-GQSM and TI-MBM techniques at all values of $T_a$, and the maximum achieved rate of TI-GQSM-MBM scheme reaches the achieved rate of GQSM-MBM scheme at the optimum value of $T_a = 126$.

*E. Maximum Likelihood (ML) Optimum Detection*

At the receiver of IM schemes, the received signal vector can be expressed as in (2), and the maximum likelihood (ML) optimum detection for this scheme is written as

$$[\widehat{\mathbf{A}}_k, \widehat{\mathbf{s}}] = \arg\min_{\mathbf{S} \in \mathbf{M}^{T_a}} \|\mathbf{y} - \mathbf{HS}\|^2, \quad (10)$$

where $\widehat{\mathbf{A}}_k$ is the index of the $k$-th antenna activation matrix that has $T_a$ columns to give the activation patterns for the $T_a$ active time slots, and $\mathbf{S}$ is the signal constellation space. The detected index $\widehat{\mathbf{A}}_k$ and the estimated symbol bits $\widehat{\mathbf{s}}$ are then used to retrieve the original transmitted data bits.

At the receiver, the imperfect channel estimation should be taken into consideration in order to be practical and to decode the transmitted information. In this paper, the estimate of $\mathbf{H}$ is denoted by $\widetilde{\mathbf{H}}$, and both are assumed to be jointly ergodic and stationary processes, and an orthogonality between the channel estimation errors (CEEs) and the channel estimate is also assumed. The estimated channel matrix $\widetilde{\mathbf{H}}$ with error is modeled as [18]

$$\widetilde{\mathbf{H}} = \mathbf{H} - \mathbf{E}, \quad (11)$$

where $\mathbf{E} \in \mathbb{C}^{TN_r \times TN_t 2^{m_{RF}}}$ represents the CEE matrix with independent and identically distributed (i.i.d.) entries that having zero mean and variance $\sigma_e^2$, $\mathbf{E} \sim \mathcal{CN}(\mathbf{0}_{N_r}, \sigma_e^2 \mathbf{I}_{N_r})$. This error variance captures the channel estimation quality, and it can be chosen appropriately depending on the channel estimation and dynamics schemes. For orthogonal pilot designs, it is assumed in this paper that the estimation error reduces linearly as the number of pilots increases [18, 19], thus the error variance $\sigma_e^2$ is kept equal to the noise variance $\sigma_n^2$.

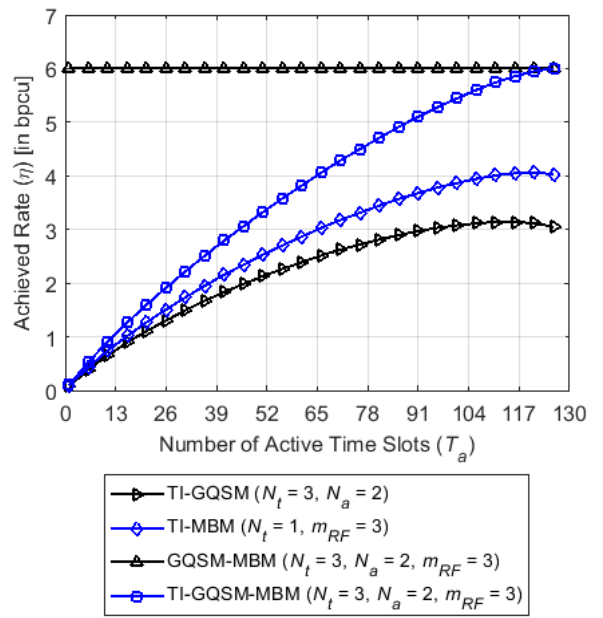

Fig. 2. The achieved rate versus $T_a$ for TI-GQSM, TI-MBM, TI-GQSM-MBM schemes with $T = 128$, and GQSM-MBM scheme with $T = T_a = 1$. All schemes are employed with $M = 2$ and $L = 1$.

Due to the sparsity of the codewords of the proposed TI-GQSM-MBM scheme, as well as the high ML receiver complexity that can be found as $\mathcal{O}\left(\frac{T^{T_a+1}(N_t^2 M 2^{m_{RF}})^{T_a} N_r}{T_a^{T_a-1}}\right)$, which is exponential in $T_a$, advanced signal processing and sparse recovery algorithms such as compressive sensing based detection methods [20] may be employed to further reduce the receiver complexity of the proposed IM schemes.

III. PERFORMANCE RESULTS AND COMPARISONS

In this section, the BER performance of the proposed TI-GQSM-MBM scheme is presented and compared to that of the relevant benchmark schemes, and all schemes are configured to achieve the same rate of 4 bpcu. Throughout the simulation, a flat Rayleigh fading channel model is assumed with $L = 1$. For all time-indexed schemes, it is assumed that $T_a = 2$ active time slots out of $T = 4$ signaling time slots.

Fig. 3 characterizes and compares the BER performance of GSM, TI-GSM and TI-GSM-MBM schemes to the performance of TI-MBM scheme that is based on indexing only time slots and RF mirrors, with assuming that a perfect channel state information (CSI) is available at the receiver. This figure shows that TI-GSM-MBM outperforms the conventional GSM, TI-GSM and TI-MBM schemes by about 10 dB, 5.5 dB and 3 dB to achieve the BER of $10^{-5}$, respectively.

In Fig. 4, the BER performance of the proposed TI-GQSM and TI-GQSM-MBM schemes is evaluated and compared to the performance of the conventional GQSM scheme and TI-MBM for the same achieved rate of 4 bpcu, with assuming that a perfect CSI is known at the receiver. It is clear from this figure that performance improvements of almost 11 dB, 7 dB and 3 dB are achieved by the proposed TI-GQSM-MBM scheme over the conventional GQSM, TI-GQSM and TI-MBM schemes at the BER of $10^{-5}$, respectively. This shows that the improvement achieved in multidimensional GQSM scheme is higher than that of its GSM counterpart in which indexing is done in time slots, antennas and RF mirrors.

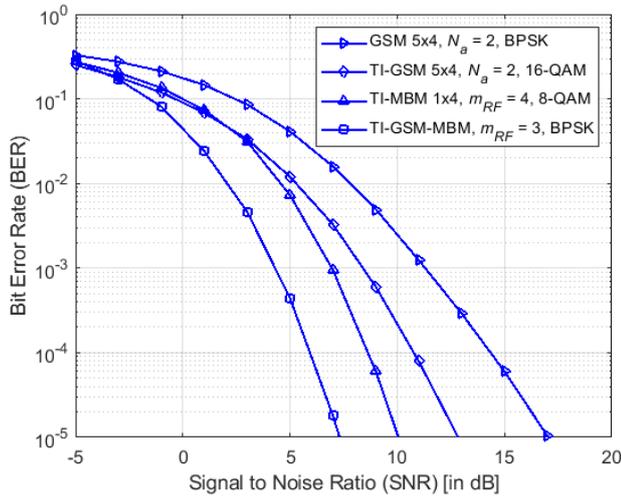

Fig. 3. BER performance of GSM, TI-GSM, TI-MBM and TI-GSM-MBM schemes for 4 bpcu using ML detection, and $L = 1$. All GSM schemes are equipped with $N_t = 5$, $N_a = 2$ and $N_r = 4$ antennas. It is assumed that $T_a = 2$ active time slots out of $T = 4$ signaling time slots in time-indexed schemes.

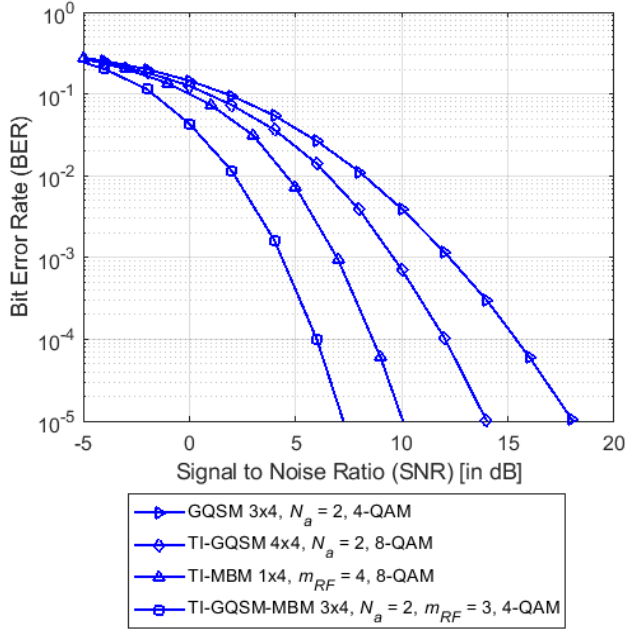

Fig. 4. BER performance of GQSM, TI-GQSM, TI-MBM and TI-GQSM-MBM schemes for 4 bpcu using ML detection, and $L = 1$. It is assumed that $T_a = 2$ active time slots out of $T = 4$ signaling time slots in time-indexed schemes.

Moreover, Figs. 3 and 4 show that the proposed TI-GSM and TI-GQSM techniques outperform the conventional GSM and GQSM schemes by about 4.5 dB and 4 dB at the BER of $10^{-5}$, respectively. It can be also seen from these two figures that the same configuration is used in both GSM and TI-GSM-MBM schemes, as well as in GQSM and TI-GQSM-MBM techniques, and significant improvements are achieved when indexing is performed in time and channel domains besides indexing in the spatial domain.

In order to explore the effect of CEEs on the BER performance of the proposed schemes, Fig. 5 shows the performance of GSM, TI-GSM and TI-GSM-MBM schemes with assuming that an imperfect CSI is available at the receiver. It is obvious from this comparison that performance degradations of about 3 dB, 4 dB and 6.5 dB are shown at the BER of $10^{-5}$ in the conventional GSM, TI-GSM and TI-

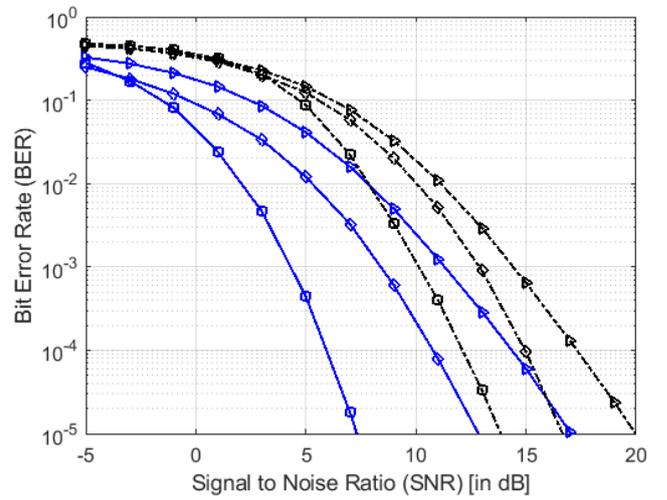

Fig. 5. BER performance of GSM, TI-GSM and TI-GSM-MBM schemes with CEEs for 4 bpcu using ML detection, and $L = 1$. All schemes are equipped with $N_t = 5$, $N_a = 2$ and $N_r = 4$ antennas. It is assumed that $T_a = 2$ active time slots out of $T = 4$ signaling time slots in TI-GSM and TI-GSM-MBM schemes.

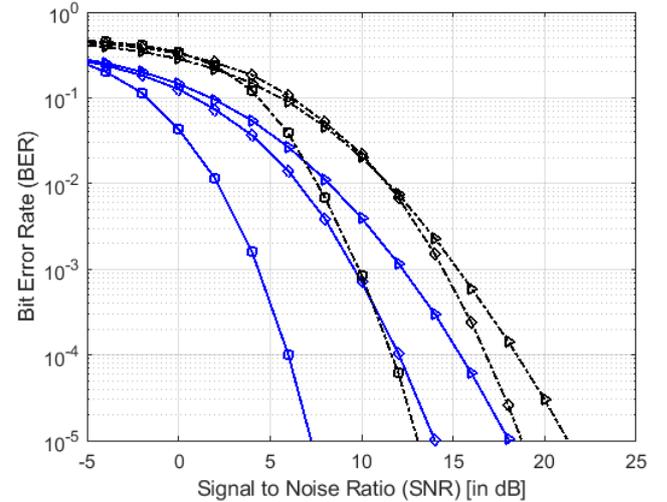

Fig. 6. BER performance of GQSM, TI-GQSM and TI-GQSM-MBM schemes with CEEs for 4 bpcu using ML detection, and $L = 1$. It is assumed that $T_a = 2$ active time slots out of $T = 4$ signaling time slots in TI-GQSM and TI-GQSM-MBM schemes, and $N_a = 2$ active antennas in all schemes.

GSM-MBM schemes, respectively, with the presence of CEEs as compared to the same schemes with perfect CSI. For multidimensional GQSM schemes, the achievable BER performance of GQSM, TI-GQSM and TI-GQSM-MBM schemes is shown in Fig. 6, where an imperfect CSI is assumed at the receiver. This figure shows that performance degradations of almost 3 dB, 4.5 dB and 6 dB are observed at the BER of $10^{-5}$ in the conventional GQSM, TI-GQSM and TI-GQSM-MBM schemes, respectively, with the presence of CEEs over the same schemes with perfect channel knowledge at the receiver. It is also noticeable from Figs. 5 and 6 that both TI-GSM-MBM and TI-GQSM-MBM schemes provide nearly the same performance when perfect CSI is assumed, with the advantage of reduced number of transmit antennas in GQSM schemes. On the other hand, it is shown that TI-GQSM-MBM outperforms TI-GSM-MBM scheme by about 0.5 dB at the BER of $10^{-5}$ when taking the CEEs into account. In addition, it is clear that the performance degradation with the presence of CEEs in the proposed multidimensional schemes is higher than that of time-indexed and conventional GSM and GQSM systems. Thus, this shows that the higher the number of transmission entities, the higher the performance drop that occurs with the presence of CEEs as compared to the same schemes with perfect channel knowledge at the receiver.

Finally, the effect of varying the number of receive antennas on the BER performance of the three proposed schemes, namely, TI-GQSM, GQSM-MBM and TI-GQSM-MBM, is evaluated in Fig. 7 at the SNR value of 6 dB with the presence of CEEs. It can be clearly seen from this figure that TI-GQSM-MBM scheme requires a smaller number of receive antennas than TI-GQSM and GQSM-MBM to achieve the same BER performance and the same rate (4 bpcu). For example, to achieve the BER of $10^{-3}$, only 8 and 14 receive antennas are required in TI-GQSM-MBM and GQSM-MBM schemes, respectively, while TI-GQSM scheme requires more than 30 receive antennas to obtain the same performance. Thus, this shows that indexing multiple transmission entities reduces the required number of receive antennas.

## IV. CONCLUSION

In this paper, a multidimensional generalized quadrature index modulation scheme in which combinations of antennas, time slots and RF mirrors are indexed simultaneously is proposed. The proposed scheme takes the advantage of QSM, GSM and MBM approaches such as the high spectral efficiency and the simple transceiver structure by combining these schemes in a joint transmission mechanism. The simulation results demonstrated that indexing multiple transmission entities is a beneficial approach, where the BER performance improves as the number of transmission entities increases. Therefore, the proposed schemes are very efficient and highly suitable techniques for large-scale MIMO systems and 5G wireless communications.

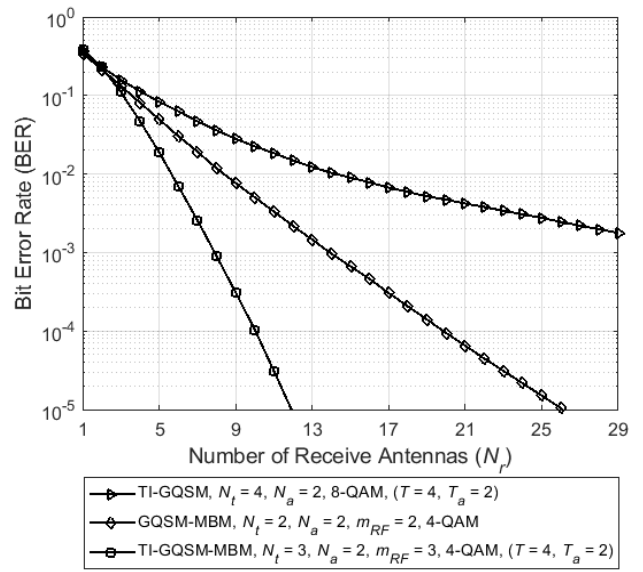

Fig. 7. BER performance of TI-GQSM, GQSM-MBM and TI-GQSM-MBM schemes in the presence of CEEs versus the number of receive antennas for 4 bpcu at the SNR value of 6 dB and $L=1$.

## REFERENCES

[1] R. Mesleh, H. Haas, S. Sinanovic, C. W. Ahn and S. Yun, "Spatial modulation," IEEE Trans. Veh. Technol., vol. 57, no. 4, pp. 2228-2241, Jul. 2008.

[2] S. Jacob, T. L. Narasimhan and A. Chockalingam, "Space-time index modulation," in Proc. IEEE Wireless Commun. Netw. Conf. (WCNC), San Francisco, CA, USA, Mar. 2017.

[3] E. Basar, U. Aygölü, E. Panayirci and H. V. Poor, "Orthogonal frequency division multiplexing with index modulation," IEEE Trans. Signal Process., vol. 61, no. 22, pp. 5536-5549, Nov. 2013.

[4] A. K. Khandani, "Media-based modulation: A new approach to wireless transmission," in Proc. IEEE Int. Symp. Inf. Theory (ISIT), Istanbul, Turkey, pp. 3050-3054, Jul. 2013.

[5] A. K. Khandani, "Media-based modulation: Converting static Rayleigh fading to AWGN," in Proc. IEEE Int. Symp. Inf. Theory (ISIT), Honolulu, HI, USA, pp. 1549-1553, Jun. 2014.

[6] S. D. Tusha, A. Tusha, E. Başar and H. Arslan, "Multidimensional index modulation for 5G and beyond wireless networks," in Proc. IEEE, vol. 109, no. 2, pp. 170-199, Feb. 2021.

[7] A. Younis, N. Serafimovski, R. Mesleh and H. Haas, "Generalised spatial modulation," in Proc. Conf. Rec. 44th Asilomar Signals Syst. Comput., Pacific Grove, CA, USA, Nov. 2010.

[8] R. Mesleh, S. S. Ikki and H. M. Aggoune, "Quadrature spatial modulation," IEEE Trans. Veh. Technol., vol. 64, no. 6, pp. 2738-2742, June 2015.

[9] F. R. Castillo-Soria, J. Cortez-Gonzá́lez, R. Ramirez-Gutierrez, F. M. Maciel-Barboza and L. Soriano-Equigua, "Generalized quadrature spatial modulation scheme using antenna grouping," ETRI Journal, vol. 39, no. 5, pp. 707-717, Oct. 2017.

[10] R. Mesleh, O. Hiari and A. Younis, "Generalized space modulation techniques: hardware design and considerations," Physical Communication, vol. 26, pp. 87-95, 2018.

[11] B. Shamasundar, S. Bhat, S. Jacob and A. Chockalingam, "Multidimensional index modulation in wireless communications," IEEE Access, vol. 6, pp. 589-604, 2017.

[12] E. Basar and I. Altunbas, "Space-time channel modulation," IEEE Trans. Veh. Technol., vol. 66, no. 8, pp. 7609-7614, Feb. 2017.

[13] N. Pillay and H. Xu, "Uncoded space-time labeling diversity—application of media-based modulation with RF mirrors," IEEE Commun. Lett., vol. 22, no. 2, pp. 272-275, Feb. 2018.

[14] Y. Naresh and A. Chockalingam, "On media-based modulation using RF mirrors," IEEE Trans. Veh. Technol., vol. 66, no. 6, pp. 4967-4983, June 2017.

[15] I. Yildirim, E. Basar and I. Altunbas, "Quadrature channel modulation," IEEE Wireless Commun. Lett., vol. 6, no. 6, pp. 790-793, Dec. 2017.

[16] B. Shamasundar, S. Jacob and A. Chockalingam, "Time-indexed media-based modulation," in Proc. IEEE 85th Veh. Technol. Conf. (VTC Spring), Sydney, NSW, Australia, Jun. 2017.

[17] T. M. Cover and J. A. Thomas, Elements of Information Theory. 2nd edition, Hoboken, NJ, USA: Wiley, July 2006.

[18] E. Basar, U. Aygolu, E. Panayirci and H. Vincent Poor, "Performance of spatial modulation in the presence of channel estimation errors," IEEE Commun. Lett., vol. 16, no. 2, pp. 176-179, Feb. 2012.

[19] J. Wu and C. Xiao, "Optimal diversity combining based on linear estimation of Rician fading channels," IEEE Trans. Commun., vol. 56, no. 10, pp. 1612-1615, Oct. 2008.

[20] D. L. Donoho, "Compressed sensing," IEEE Trans. Inf. Theory, vol. 52, no. 4, pp. 1289-1306, Apr. 2006.